\documentclass[letterpaper,twocolumn,reprint,showpacs,prb,amsmath,amssymb,floatfix,eqsecnum]{revtex4-1}
\usepackage[T1]{fontenc}
\usepackage[latin9]{inputenc}
\usepackage{color}
\usepackage{bm}
\usepackage{amsmath}
\usepackage{amssymb}
\usepackage{graphicx}
\usepackage{wasysym}
\usepackage{esint}
\usepackage[unicode=true,pdfusetitle,
 bookmarks=true,bookmarksnumbered=false,bookmarksopen=false,
 breaklinks=false,pdfborder={0 0 0},backref=false,colorlinks=false]
 {hyperref}

\makeatletter

\pdfpageheight\paperheight
\pdfpagewidth\paperwidth

\usepackage[toc]{appendix}

\makeatother

\begin{document}

\title{Hard-core boson approach to the spin-1/2 triangular-lattice antiferromagnet
Cs$_{2}$CuCl$_{4}$ at finite temperatures in magnetic fields higher
than the saturation field }

\author{Simon Streib}

\email{streib@itp.uni-frankfurt.de}

\affiliation{Institut für Theoretische Physik, Universität Frankfurt, Max-von-Laue
Strasse 1, 60438 Frankfurt, Germany}

\author{Peter Kopietz}

\affiliation{Institut für Theoretische Physik, Universität Frankfurt, Max-von-Laue
Strasse 1, 60438 Frankfurt, Germany}

\date{September 30, 2015}
\begin{abstract}
We study the high magnetic field regime of the antiferromagnetic insulator
Cs$_{2}$CuCl$_{4}$ by expressing the spin-1/2 operators in the relevant
Heisenberg model in terms of hard-core bosons and implementing the
hard-core constraint via an infinite on-site interaction. We focus
on the case where the external magnetic field exceeds the saturation
field $B_{c}\approx8.5\;\mathrm{T}$ and is oriented along the crystallographic
$a$ axis perpendicular to the lattice plane. Because in this case
the excited states are separated by an energy gap from the ground
state, we may use the self-consistent ladder approximation to take
the strong correlations due to the hard-core constraint into account.
In Cs$_{2}$CuCl$_{4}$ there are additional interactions besides
the hard-core interaction which we treat in self-consistent Hartree-Fock
approximation. We calculate the spectral function of the hard-core
bosons from which we obtain the in-plane components of the dynamic
structure factor, the magnetic susceptibility, and the specific heat.
Our results for the specific heat are in good agreement with the available
experimental data. We conclude that the self-consistent ladder approximation
in combination with a self-consistent Hartree-Fock decoupling of the
non-hard-core interactions gives an accurate description of the physical
properties of gapped hard-core bosons in two dimensions at finite
temperatures. 
\end{abstract}


\maketitle

\section{Introduction}

The magnetic behavior of the antiferromagnetic insulator Cs$_{2}$CuCl$_{4}$
can be described by a spin-1/2 Heisenberg model on a triangular lattice
where the interlayer coupling is much weaker than the intralayer couplings.
Due to the relatively weak exchange couplings, a field induced ferromagnetic
ground-state can be reached at fields larger than the saturation field
\textbf{$B_{c}\approx8.5\;\mathrm{T}$} where the magnetic field is
along the crystallographic $a$ axis perpendicular to the lattice
plane. This allows a precise measurement of the exchange couplings
via inelastic neutron scattering experiments where the single magnon
dispersion gives direct access to the exchange couplings.\cite{Coldea02}
Cs$_{2}$CuCl$_{4}$ has been intensively studied due to its interesting
properties, e.g., spin-liquid behavior with spinon excitations,\cite{Coldea01,Coldea03,Yunoki06,Weng06,Hayashi07,Kohno07,Kohno09,Heidarian09,Balents10,Tay10,Vachon11,Herfurth13,Tocchio14}
Bose-Einstein condensation of magnons at the quantum critical point,\cite{Coldea02,Radu05,Radu07,Kovrizhin06,Zapf14}
and a rich phase diagram for in-plane magnetic fields.\cite{Coldea01,Veillette05,Tokiwa06,Starykh10,Griset11,Chen13,Starykh15}A
diverse range of observables have been experimentally investigated:
dynamic structure factor,\cite{Coldea01,Coldea02,Coldea03} electron
spin resonance spectra,\cite{Zvyagin14} magnetic susceptibility,\cite{Tokiwa06}
magneto-caloric effect,\cite{Lang13} nuclear magnetic resonance relaxation
rate,\cite{Vachon11} specific heat,\cite{Radu05,Radu07} and ultrasound
velocity and attenuation.\cite{Sytcheva09,Kreisel11,Streib15}

In this work, we consider the case of a large magnetic field $B>B_{c}$
along the $a$ axis, where the magnon excitations are gapped and the
ground state is the fully magnetized ferromagnet. Our goal is to describe
the thermal excitations above the ground state and to compare with
experimental results for the specific heat.\cite{Radu05,Radu07} We
base our theoretical approach on a mapping of the spin-1/2 operators
to hard-core bosons.\cite{Matsubara56,Batyev84} For magnetic fields
$B>B_{c}$ and low temperatures, we then have a dilute gas of gapped
hard-core bosons where the ladder approximation captures the leading
order low-temperature contributions to the self-energy. Although the
ladder approximation has been extensively applied to the Bose-condensed
phase of dilute gases of hard-core bosons (see Ref.~{[}\onlinecite{Zapf14}{]}
and references therein), some subtleties related to the hard-core
limit in the gapped phase have only recently been discussed by Fauseweh,
Stolze, and Uhrig (FSU).\cite{Fauseweh14,Fauseweh15} Benchmarking
the ladder approximation for an exactly solvable one-dimensional model
of hard-core bosons, FSU found that the ladder approximation indeed
reproduces the correct low-temperature behavior and that a self-consistent
ladder approximation even extends the applicability to arbitrarily
high temperatures.\cite{Fauseweh14} In this work, we apply the self-consistent
ladder approximation to the relevant two-dimensional model for Cs$_{2}$CuCl$_{4}$.
For a realistic description of this material we have to include additional
interactions apart from the infinite on-site interaction describing
the hard-core constraint. To further explore the range of validity
of the self-consistent ladder approximation, we have also applied
this method to the exactly solvable one-dimensional $XY$ model; extending
the analysis of FSU,\cite{Fauseweh14} we have examined the breakdown
of the self-consistent ladder approximation in the vicinity of the
quantum critical point of this model.

The rest of this work is organized as follows. In the next section,
we introduce the relevant spin model for Cs$_{2}$CuCl$_{4}$ and
describe the mapping of this model to an effective hard-core boson
model. Then, in Sec.~\ref{sec:theoretical-approach}, we describe
our theoretical approach based on the self-consistent ladder approximation
for the hard-core interaction and a self-consistent Hartree-Fock decoupling
for the remaining non-hard-core interactions. In Sec.~\ref{sec:numerical-Results XY},
we investigate the breakdown of the ladder approximation near the
quantum critical point for the exactly solvable one-dimensional $XY$
model and in Sec.~\ref{sec:numerical-Results Cs2CuCl4} we present
our numerical results for Cs$_{2}$CuCl$_{4}$, which we compare with
experimental data for the specific heat. Finally, in Sec.~\ref{sec:summary}
we summarize our main results. In three appendices we give additional
technical details of our calculations.

\section{Hard-core boson model for $\mathbf{Cs_{2}CuCl_{4}}$ \label{sec:model}}

It has been established that the magnetic behavior of Cs$_{2}$CuCl$_{4}$
can be described by the following two-dimensional antiferromagnetic
spin-$1/2$ Heisenberg model in an external magnetic field along the
crystallographic $a$ axis,\cite{Coldea02} 
\begin{figure}
\begin{centering}
\includegraphics{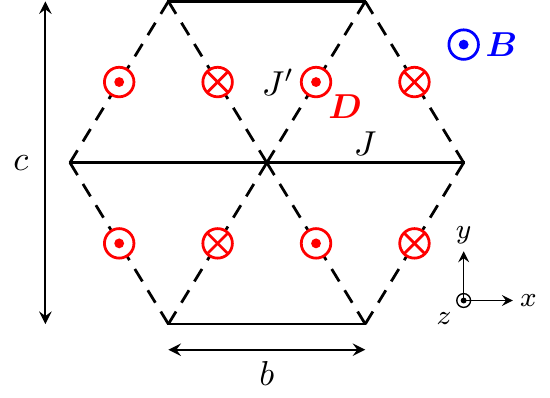}
\par\end{centering}

\caption{(Color online) Part of the anisotropic triangular lattice formed by
the spins of Cs$_{2}$CuCl$_{4}$. The stronger exchange coupling
$J$ connects nearest-neighbor spins along the crystallographic $b$
axis, while the weaker exchange coupling $J'$ connects nearest-neighbor
spins along the diagonals. There are also weak Dzyaloshinskii-Moriya
interactions $\bm{D}=\pm D\hat{\bm{z}}$ connecting neighboring spins
along the diagonals where the direction of $\bm{D}$ is indicated
by \textcolor{red}{$\bm{\odot}$} for $+\hat{\bm{z}}$ and\textcolor{red}{{}
$\bm{\otimes}$} for $-\hat{\bm{z}}$. We consider only the case where
the magnetic field $\bm{B}=B\hat{\bm{z}}$ is along the $a$ axis
perpendicular to the plane of the lattice.\label{fig:lattice}}
\end{figure}

\begin{equation}
\ensuremath{\mathcal{H}=\frac{1}{2}\sum_{ij}\left[J_{ij}\bm{S}_{i}\cdot\bm{S}_{j}+\bm{D}_{ij}\cdot(\bm{S}_{i}\times\bm{S}_{j})\right]-h\sum_{i}S_{i}^{z}},\label{eq:hamiltonian}
\end{equation}
where the summations run over all $N$ lattice sites, $h=g\mu_{B}B$
is the Zeeman energy associated with an external magnetic field $\bm{B}=B\hat{\bm{z}}$,
and $g=2.19(1)$ is the effective $g$-factor.\cite{Coldea02} The
spin-$1/2$ operators $\bm{S}_{i}=\bm{S}(\bm{R}_{i})$ are located
at the lattice sites $\bm{R}_{i}$ of an anisotropic triangular lattice
with lattice constants $b$ and $c$, as shown in Fig.~\ref{fig:lattice}.
The exchange couplings $J_{ij}=J(\bm{R}_{i}-\bm{R}_{j})$ connect
nearest neighbors along the crystallographic $b$ axis and along the
diagonals with $J(\pm\bm{\delta}_{1})=J$ and $J(\pm\bm{\delta}_{2})=J(\pm\bm{\delta}_{3})=J'$,
where the three elementary lattice vectors are 
\begin{equation}
\bm{\delta}_{1}=b\hat{\bm{x}},\;\bm{\delta}_{2}=-\frac{b}{2}\hat{\bm{x}}+\frac{c}{2}\hat{\bm{y}},\;\bm{\delta}_{3}=-\frac{b}{2}\hat{\bm{x}}-\frac{c}{2}\hat{\bm{y}}.
\end{equation}
Here, $\hat{\bm{x}}$, $\hat{\bm{y}}$, and $\hat{\bm{z}}$ are the
unit vectors of our Cartesian coordinate system. Due to the fact that
inversion symmetry is broken for Cs$_{2}$CuCl$_{4}$, there are also
Dzyaloshinskii-Moriya (DM) interactions $\bm{D}_{ij}=D(\bm{R}_{i}-\bm{R}_{j})\hat{\bm{z}}$
connecting neighboring spins along the diagonals with $D(\pm\bm{\delta}_{2})=D(\pm\bm{\delta}_{3})=\mp D$.
The precise form of the Hamiltonian (\ref{eq:hamiltonian}) and the
values of the interaction constants have been measured by inelastic
neutron scattering experiments in magnetic fields higher than the
saturation field $B_{c}=8.44(1)\;\mathrm{T}$. The accepted values
are:\cite{Coldea02} $J=0.374(5)\;\mathrm{meV}=4.34(6)\;\mathrm{K}$,
$J'/J=0.34(3)$, and $D/J=0.053(5)$. There is also a weak interlayer
coupling $J''/J=0.045(5)$, which we neglect because it is only important
at very low temperatures $T\lesssim0.1\;\mathrm{K}$ and in the antiferromagnetically
ordered phase in magnetic fields $B<B_{c}$ which we do not consider
here. Recently, additional DM interactions, including in-plane components,
have been measured via electron spin resonance experiments.\cite{Zvyagin14}
We neglect these additional DM interactions because they are mainly
important for in-plane magnetic fields.\cite{Starykh10} Furthermore,
our theoretical approach relies on the $U(1)$ symmetry due to the
spin-rotational invariance with respect to the $z$ axis, which would
be broken by in-plane DM interactions.

In this work, we will use the hard-core boson representation of the
spin-$1/2$ operators.\cite{Matsubara56,Batyev84} Recall that the
spin-1/2 operators fulfill the commutation relations 
\begin{equation}
\left[S_{i}^{+},S_{j}^{-}\right]=2\delta_{ij}S_{i}^{z},\;\;\;\left[S_{i}^{\pm},S_{j}^{z}\right]=\mp\delta_{ij}S_{i}^{\pm},
\end{equation}
where $S_{i}^{\pm}=S_{i}^{x}\pm iS_{i}^{y}$ and $\bm{S}_{i}^{2}=3/4$.
Additionally, the spin-1/2 operators obey an on-site exclusion principle,\cite{Wang92}
\begin{equation}
S_{i}^{+}S_{i}^{-}+S_{i}^{-}S_{i}^{+}=1,\;\;\;(S_{i}^{+})^{2}=(S_{i}^{-})^{2}=0.
\end{equation}
To realize these relations, we can express the spin operators in terms
of hard-core boson creation and annihilation operators,

\begin{equation}
S_{i}^{+}=b_{i},\quad S_{i}^{-}=b_{i}^{\dagger},\quad S_{i}^{z}=1/2-b_{i}^{\dagger}b_{i},\label{eq:hard-core boson representation}
\end{equation}
where the hard-core boson operators satisfy the commutation relation
\begin{equation}
\left[b_{i},b_{j}^{\dagger}\right]=\delta_{ij}\left(1-2b_{i}^{\dagger}b_{i}\right),\label{eq:hard-core boson commutator}
\end{equation}
and the occupation number per site is restricted to $\hat{n}_{i}=0$
or $1$, with $\hat{n}_{i}=b_{i}^{\dagger}b_{i}$. The hard-core boson constraint and the commutation relation
(\ref{eq:hard-core boson commutator}) can be realized by treating
the hard-core bosons as canonical bosons with an infinite on-site
repulsion, 
\begin{equation}
\mathcal{H}_{U}=\frac{U}{2}\sum_{i}b_{i}^{\dagger}b_{i}^{\dagger}b_{i}b_{i},\,\ensuremath{\,\mathrm{with}\,\,U\to\infty}.
\end{equation}
Note that the magnon excitations of the underlying spin system correspond
to hard-core boson excitations.

\begin{figure}
\begin{centering}
\includegraphics{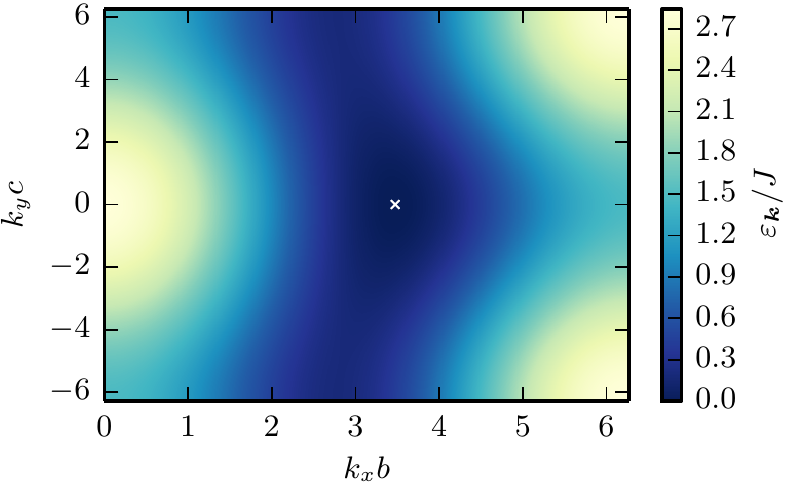}
\par\end{centering}

\caption{(Color online) Contour plot of the energy dispersion $\varepsilon_{\bm{k}}$
defined in Eq.~(\ref{eq:e_k}). The white cross marks the minimum
of $\varepsilon_{\bm{k}}$ at $\bm{Q}\approx(3.474/b,0)$.\label{fig: contour of epsilon}}
\end{figure}

Using Eq.~(\ref{eq:hard-core boson representation}) to express the
spin operators in our Hamiltonian (\ref{eq:hamiltonian}) in terms
of hard-core bosons, we obtain the following hard-core boson Hamiltonian,
\begin{equation}
\mathcal{H}=\sum_{\bm{k}}\xi_{\bm{k}}b_{\bm{k}}^{\dagger}b_{\bm{k}}+\frac{1}{2N}\sum_{\bm{k},\bm{k}',\bm{q}}(J_{\bm{q}}+U)b_{\bm{k}+\bm{q}}^{\dagger}b_{\bm{k}'-\bm{q}}^{\dagger}b_{\bm{k}'}b_{\bm{k}}+E_{0},\label{eq:hard-core boson Hamiltonian}
\end{equation}
where we have Fourier transformed the hard-core boson creation and
annihilation operators, 
\begin{equation}
b_{\bm{k}}=\frac{1}{\sqrt{N}}\sum_{i}b_{i}e^{-i\bm{k}\cdot\bm{R}_{i}},\;\;\;b_{\bm{k}}^{\dagger}=\frac{1}{\sqrt{N}}\sum_{i}b_{i}^{\dagger}e^{i\bm{k}\cdot\bm{R}_{i}}.
\end{equation}
In the following, we will neglect the unimportant constant energy
term 
\begin{equation}
E_{0}=N\left(\frac{J_{0}}{8}-\frac{h}{2}\right).
\end{equation}
The excitation energy $\xi_{\bm{k}}$ in the quadratic part of the
Hamiltonian can be written as 
\begin{equation}
\xi_{\bm{k}}=\varepsilon_{\bm{k}}-\mu,\label{eq:bare excitation energy}
\end{equation}
where we have introduced the chemical potential 
\begin{equation}
\mu=h_{c}-h,
\end{equation}
and the energy dispersion 
\begin{equation}
\varepsilon_{\bm{k}}=\frac{1}{2}\left(J_{\bm{k}}^{D}-J_{\bm{Q}}^{D}\right).\label{eq:e_k}
\end{equation}
Here 
\begin{equation}
J_{\bm{k}}^{D}=J_{\bm{k}}-iD_{\bm{k}},
\end{equation}
where the Fourier transforms of the exchange and Dzyaloshinskii-Moriya
interactions are 
\begin{eqnarray}
J_{\bm{k}} & = & \sum_{\bm{R}}J(\bm{R})e^{-i\bm{k}\cdot\bm{R}}\nonumber \\
 & = & 2J\cos(k_{x}b)+4J'\cos\left(\frac{k_{x}b}{2}\right)\cos\left(\frac{k_{y}c}{2}\right),\nonumber \\
\\
D_{\bm{k}} & = & \sum_{\bm{R}}D(\bm{R})e^{-i\bm{k}\cdot\bm{R}}\nonumber \\
 & = & -4iD\sin\left(\frac{k_{x}b}{2}\right)\cos\left(\frac{k_{y}c}{2}\right).
\end{eqnarray}
In Eq.~(\ref{eq:e_k}), $J_{\bm{Q}}^{D}\approx-2.325\;J$ is the
absolute minimum of $J_{\bm{k}}^{D}$ at $\bm{Q}\approx(3.474/b,0)$.
Finally, the saturation field is given by 
\begin{equation}
B_{c}=\frac{h_{c}}{g\mu_{B}}=\frac{1}{2g\mu_{B}}\left(J_{0}^{D}-J_{\bm{Q}}^{D}\right)\approx8.4\;\mathrm{T}.
\end{equation}
A contour plot of $\varepsilon_{\bm{k}}$ is shown in Fig.~\ref{fig: contour of epsilon}.
In the following, we will use the direct experimental value of the
saturation field $B_{c}=8.44(1)\;\mathrm{T}$ instead of $B_{c}\approx8.4\;\mathrm{T}$
because the experimental value is more accurate than a calculation
via the Hamiltonian (\ref{eq:hard-core boson Hamiltonian}). The reason
is that the interaction constants in Eq.~(\ref{eq:hard-core boson Hamiltonian})
have some experimental uncertainty and we have also neglected the
interlayer coupling $J''$; including $J''$ in the calculation would
result in $B_{c}\approx8.5\;\mathrm{T}$.\cite{Radu05,Kovrizhin06}
The value of the saturation field $B_{c}$ is important because, for
a given magnetic field, it determines the energy gap 
\begin{equation}
\Delta=-\mu=h-h_{c}.
\end{equation}
We note that a small change of $B_{c}$ by $0.04\;\mathrm{T}$ changes
the gap by about $0.014\;J$, which is only significant close to the
quantum critical point.

\section{Implementing the self-consistent ladder approximation\label{sec:theoretical-approach}}

In this section, we explain our theoretical approach to the hard-core
boson Hamiltonian (\ref{eq:hard-core boson Hamiltonian}). The central
problem is how to deal with the interaction part of the Hamiltonian,

\begin{equation}
\mathcal{H}_{\mathrm{int}}=\frac{1}{2N}\sum_{\bm{k},\bm{k}',\bm{q}}(J_{\bm{q}}+U)b_{\bm{k}+\bm{q}}^{\dagger}b_{\bm{k}'-\bm{q}}^{\dagger}b_{\bm{k}'}b_{\bm{k}},\label{eq:interaction part}
\end{equation}
containing the exchange interaction $J_{\bm{q}}$ and the infinite
hard-core interaction $U\to\infty$. We will deal with both interactions
using different methods: for the $J_{\bm{q}}$ part we use a self-consistent
Hartree-Fock decoupling, while for the hard-core interaction $U$
we use the self-consistent ladder approximation.\cite{Fauseweh14}
This is necessary because the $J_{\bm{q}}$ interaction cannot be
easily included in the self-consistent ladder approximation, as this
would not allow a direct solution for the effective interaction $\Gamma$
from the Bethe-Salpeter equation which would significantly complicate
matters, especially regarding the limit $U\to\infty$.

\subsection{Hartree-Fock decoupling}

We approximate the $J_{\bm{q}}$ interaction term in Eq.~(\ref{eq:interaction part})
using a self-consistent Hartree-Fock decoupling. Therefore, we write
this term in real space,

\begin{equation}
\frac{1}{2N}\sum_{\bm{k},\bm{k}',\bm{q}}J_{\bm{q}}b_{\bm{k}+\bm{q}}^{\dagger}b_{\bm{k}'-\bm{q}}^{\dagger}b_{\bm{k}'}b_{\bm{k}}=\frac{1}{2}\sum_{i,j}J_{ij}b_{i}^{\dagger}b_{i}b_{j}^{\dagger}b_{j},\label{eq:J_q term}
\end{equation}
and then we apply the usual Hartree-Fock decoupling, 
\begin{eqnarray}
b_{i}^{\dagger}b_{i}b_{j}^{\dagger}b_{j} & \approx & n_{i}b_{j}^{\dagger}b_{j}+n_{j}b_{i}^{\dagger}b_{i}-n_{i}n_{j}\nonumber \\
 & + & \tau_{ji}b_{i}^{\dagger}b_{j}+\tau_{ij}b_{j}^{\dagger}b_{i}-\tau_{ij}\tau_{ji},\label{eq:hartree-fock approximation}
\end{eqnarray}
giving 
\begin{equation}
\frac{1}{2}\sum_{i,j}J_{ij}b_{i}^{\dagger}b_{i}b_{j}^{\dagger}b_{j}\approx\sum_{i,j}J_{ij}\left(n_{j}b_{i}^{\dagger}b_{i}+\tau_{ji}b_{i}^{\dagger}b_{j}\right)+E_{\mathrm{MF}},\label{eq:hartree-fock interaction}
\end{equation}
where the Hartree-Fock parameters are given by 
\begin{equation}
n_{i}=\left\langle b_{i}^{\dagger}b_{i}\right\rangle ,\;\;\;\tau_{ij}=\left\langle b_{i}^{\dagger}b_{j}\right\rangle ,
\end{equation}
and the constant energy term is 
\begin{equation}
E_{\mathrm{MF}}=-\frac{1}{2}\sum_{i,j}J_{ij}\left(n_{i}n_{j}+\tau_{ij}\tau_{ji}\right).
\end{equation}
Due to translational invariance, we have 
\begin{equation}
n_{i}=n,\;\;\;\tau_{ij}=\tau(\bm{R}_{i}-\bm{R}_{j}).
\end{equation}
Because there is no inversion symmetry, $\tau(\bm{R})$ is a complex
number satisfying $\tau^{*}(\bm{R})=\tau(-\bm{R})$. Since the exchange
coupling $J(\bm{R})$ is only non-zero for $\bm{R}=\pm\bm{\delta}_{i}$,
there are three complex Hartree-Fock parameters related to $\tau(\bm{R})$,
\begin{equation}
\tau_{1}=\tau(\bm{\delta}_{1}),\;\;\;\tau_{2}=\tau(\bm{\delta}_{2}),\;\;\;\tau_{3}=\tau(\bm{\delta}_{3}).
\end{equation}
However, the Hamiltonian (\ref{eq:hard-core boson Hamiltonian}) is
invariant under the transformation $k_{y}\to-k_{y}$ and therefore
$\tau_{2}=\tau_{3}$. Transforming Eq.~(\ref{eq:hartree-fock interaction})
back to momentum space, we get 
\begin{equation}
\frac{1}{2N}\sum_{\bm{k},\bm{k}',\bm{q}}J_{\bm{q}}b_{\bm{k}+\bm{q}}^{\dagger}b_{\bm{k}'-\bm{q}}^{\dagger}b_{\bm{k}'}b_{\bm{k}}\approx\sum_{\bm{k}}\left(J_{\bm{k}}^{\tau}+nJ_{0}\right)b_{\bm{k}}^{\dagger}b_{\bm{k}}+E_{\mathrm{MF}},
\end{equation}
where
\begin{eqnarray}
J_{\bm{k}}^{\tau} & = & 2J\mathrm{Re}\left(\tau_{1}e^{i\bm{k}\cdot\bm{\delta}_{1}}\right)+2J'\mathrm{Re}\left(\tau_{2}e^{i\bm{k}\cdot\bm{\delta}_{2}}+\tau_{3}e^{i\bm{k}\cdot\bm{\delta}_{3}}\right),\nonumber \\
\end{eqnarray}
\begin{equation}
E_{\mathrm{MF}}=-N\left[\frac{J_{0}}{2}n^{2}+J|\tau_{1}|^{2}+J'\left(|\tau_{2}|^{2}+|\tau_{3}|^{2}\right)\right].
\end{equation}
The Hartree-Fock approximation gives a constant energy shift $E_{\mathrm{MF}}$,
which depends on the magnetic field and the temperature; moreover,
the Hartree-Fock approximation leads to a renormalization of single-particle
excitation energies $\xi_{\bm{k}}\to\tilde{\xi}_{\bm{k}}$, where
the renormalized excitation energies are 
\begin{equation}
\tilde{\xi}_{\bm{k}}=\varepsilon_{\bm{k}}-\mu+J_{\bm{k}}^{\tau}+nJ_{0}.
\end{equation}
The self-consistency equations for the Hartree-Fock parameters are
given by\begin{subequations}

\begin{eqnarray}
\tau_{i} & = & \tau(\bm{\delta}_{i})=\frac{1}{N}\sum_{\bm{k}}n_{\bm{k}}e^{-i\bm{k}\cdot\bm{\delta}_{i}},\label{eq:tau}\\
n & = & \frac{1}{N}\sum_{\bm{k}}n_{\bm{k}},\label{eq:n}
\end{eqnarray}
\end{subequations}where the occupation number of a state with momentum
$\bm{k}$ is given by 
\begin{equation}
n_{\bm{k}}=\left\langle b_{\bm{k}}^{\dagger}b_{\bm{k}}\right\rangle .
\end{equation}
If we neglect the hard-core interaction, we simply obtain the Bose-Einstein
distribution, 
\begin{equation}
n_{\bm{k}}^{\mathrm{HF}}=\frac{1}{e^{\beta\tilde{\xi}_{\bm{k}}}-1},
\end{equation}
where $\beta=1/T$ is the inverse temperature and the renormalized
excitation energy $\tilde{\xi}_{\bm{k}}$ can be obtained in a straightforward
way by solving the self-consistency equations for the Hartree-Fock
parameters for $U=0$. Neglecting the hard-core interaction is possible
only for small temperatures $T\ll J$ when the bosons are so dilute
that the hard-core interaction does not contribute significantly.

\subsection{Self-consistent ladder approximation}

After the Hartree-Fock decoupling of the $J_{\bm{q}}$ interaction,
we obtain a Hamiltonian where the only remaining interaction is the
infinite on-site repulsion, 
\begin{equation}
\mathcal{H}=\sum_{\bm{k}}\tilde{\xi}_{\bm{k}}b_{\bm{k}}^{\dagger}b_{\bm{k}}+\frac{U}{2N}\sum_{\bm{k},\bm{k}',\bm{q}}b_{\bm{k}+\bm{q}}^{\dagger}b_{\bm{k}'-\bm{q}}^{\dagger}b_{\bm{k}'}b_{\bm{k}}+E_{\mathrm{MF}}.\label{eq:hard-core boson Hamiltonian after mean-field decoupling}
\end{equation}
We will deal with this hard-core interaction using the self-consistent
ladder approximation developed in Ref.~{[}\onlinecite{Fauseweh14}{]}.

\subsubsection{Imaginary time path integral formalism}

To derive the self-consistent ladder approximation, it is convenient
to formulate the problem in terms of an imaginary time path integral.\cite{Negele+Orland}
The Euclidean action associated with the Hamiltonian (\ref{eq:hard-core boson Hamiltonian after mean-field decoupling})
is 
\begin{eqnarray}
S[\bar{b},b] & = & -\int_{K}G_{0}^{-1}(K)\bar{b}_{K}b_{K}\nonumber \\
 &  & +\frac{U}{2}\int_{K,K',Q}\bar{b}_{K+Q}\bar{b}_{K'-Q}b_{K'}b_{K}.
\end{eqnarray}
Here, we have introduced the composite index $K=\left(\bm{k},i\omega_{k}\right)$
with the corresponding sum 
\begin{equation}
\int_{K}=\frac{1}{\beta N}\sum_{\bm{k}}\sum_{\omega_{k}},
\end{equation}
where $\omega_{k}$ are bosonic Matsubara frequencies. The complex
boson fields in imaginary time have been Fourier transformed to frequency
space as \begin{subequations} 
\begin{eqnarray}
b_{\bm{k}}(\tau) & = & \frac{1}{\beta\sqrt{N}}\sum_{\omega_{k}}e^{-i\omega_{k}\tau}b_{K},\\
\bar{b}_{\bm{k}}(\tau) & = & \frac{1}{\beta\sqrt{N}}\sum_{\omega_{k}}e^{i\omega_{k}\tau}\bar{b}_{K}.
\end{eqnarray}
\end{subequations} The Green function $G(K)$ and the corresponding
self-energy $\Sigma(K)$ are defined via the functional average 
\begin{equation}
\left\langle \bar{b}_{K}b_{K}\right\rangle =-\beta NG(K)=-\beta N\frac{1}{G_{0}^{-1}(K)-\Sigma(K)},
\end{equation}
where the bare Green function $G_{0}(K)$ is given by 
\begin{equation}
G_{0}(K)=\frac{1}{i\omega_{k}-\tilde{\xi}_{\bm{k}}}.
\end{equation}
From this path integral formalism a perturbative diagrammatic expansion
of the one-particle irreducible self-energy $\Sigma(K)$ can be obtained
in terms of the bare Green function $G_{0}(K)$ and the interaction
$U$.

\subsubsection{Self-consistent ladder approximation}

\begin{figure}
\begin{centering}
\includegraphics{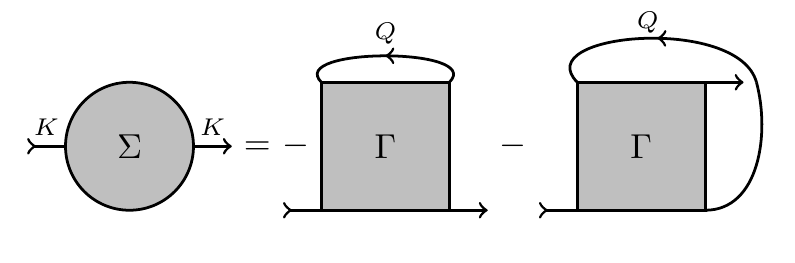} 
\par\end{centering}

\caption{Diagrammatic representation of the self-energy in terms of the effective
interaction as given in Eq.~(\ref{eq:self-energy from effective interaction}).\label{fig:self-energy}}
\end{figure}

\begin{figure}
\begin{centering}
\includegraphics{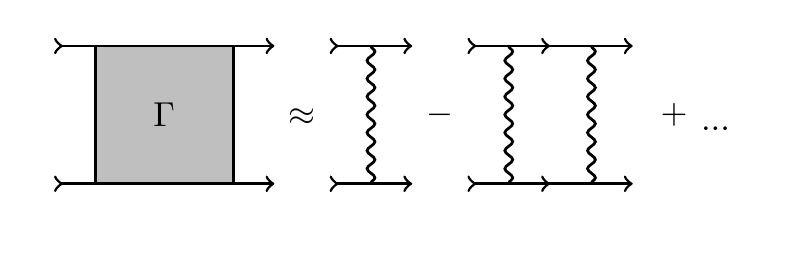} 
\par\end{centering}

\caption{Ladder approximation for the effective interaction $\Gamma$ including
all particle-particle ladder diagrams.\label{fig:ladder-diagrams}}
\end{figure}

Since we are dealing with a strictly non-perturbative problem ($U\to\infty$),
it is necessary to sum over a suitable infinite set of diagrams containing
infinite powers of $U$. Here we approximate the self-energy by summing
over all particle-particle ladder diagrams, where we express the self-energy
in terms of the effective interaction $\Gamma$, as shown in Fig.~\ref{fig:self-energy}.
The effective interaction then includes the infinite series of particle-particle
ladder diagrams indicated in Fig.~\ref{fig:ladder-diagrams}. 
Formally, this approximation is justified for $\beta\Delta\gg1$ because
the neglected diagrams are of order $\exp(-\beta\Delta)$ smaller
than the ladder diagrams. The neglected diagrams for the self-energy
include at least two lines going backwards in imaginary time while
the ladder diagrams only include a single line of this type; each
such line gives a suppression of $\exp(-\beta\Delta)$. This can be
seen by considering the bare Green function in imaginary time, 
\begin{equation}
G_{0}(\bm{k},\tau)=\begin{cases}
-\left(1+n_{B}(\tilde{\xi}_{\bm{k}})\right)e^{-\tilde{\xi}_{\bm{k}}\tau}, & \tau>0\\
-n_{B}(\tilde{\xi}_{\bm{k}})e^{-\tilde{\xi}_{\bm{k}}\tau}, & \tau<0
\end{cases},
\end{equation}
where $n_{B}(x)$ denotes the Bose function, 
\begin{equation}
n_{B}(x)=\frac{1}{e^{\beta x}-1}.
\end{equation}
We see that $n_{B}(\tilde{\xi}_{\bm{k}})\propto\exp(-\beta\Delta)$
for $\beta\Delta\gg1$ and therefore $G_{0}(\bm{k},\tau)\propto\exp(-\beta\Delta)$
for $\tau<0$.

\begin{figure}
\begin{centering}
\includegraphics{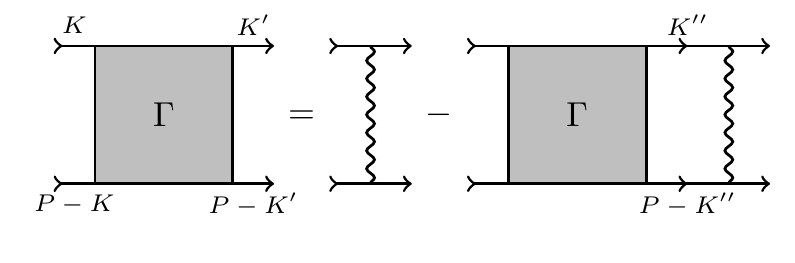} 
\par\end{centering}

\caption{Diagrammatic representation of the Bethe-Salpeter equation (\ref{eq:bethe-salpeter})
for the effective interaction $\Gamma$.\label{fig:bethe-salpeter-equation}}
\end{figure}

We can go beyond the ladder approximation and include higher order
terms by using the full Green function $G(K)$ in the diagrammatic
expansion instead of the bare Green function $G_{0}(K)$ and then
finding a self-consistent solution. In this self-consistent ladder
approximation, the effective interaction fulfills the Bethe-Salpeter
equation shown in Fig.~\ref{fig:bethe-salpeter-equation}, 

\begin{equation}
\Gamma(K',K;P)=U-U\int_{K''}\Gamma(K'',K;P)G(K'')G(P-K'').\label{eq:bethe-salpeter}
\end{equation}
Because the hard-core interaction $U$ is a constant independent of
the momentum transfer, the Bethe-Salpeter equation has the simple
solution 
\begin{equation}
\Gamma(K',K;P)=\Gamma(P)=\frac{U}{1+U\Pi(P)},\label{eq:gamma}
\end{equation}
where we have defined the particle-particle bubble 
\begin{equation}
\Pi(P)=\int_{Q}G(Q)G(P-Q).
\end{equation}
The self-energy in the self-consistent ladder approximation is given
by 
\begin{equation}
\Sigma(K)=-2\int_{Q}G(Q)\Gamma(Q+K)e^{i\omega_{q}0^{+}},\label{eq:self-energy from effective interaction}
\end{equation}
which is shown diagrammatically in Fig.~\ref{fig:self-energy}. The
convergence factor $e^{i\omega_{q}0^{+}}$ implements the correct
time ordering at the interaction vertex in the first order term in
$U$ where a propagator line starts and ends at the same vertex.\cite{Negele+Orland}
The Green function has the spectral representation 
\begin{equation}
G(K)=\int_{-\infty}^{\infty}dx\,\frac{A(\bm{k},x)}{i\omega_{k}-x},
\end{equation}
where the spectral function is given by 
\begin{eqnarray}
A(\bm{k},\omega) & = & -\frac{1}{\pi}\mathrm{Im}G(\bm{k},\omega+i0^{+})\nonumber \\
 & = & -\frac{1}{\pi}\frac{\mathrm{Im}\Sigma^{R}(\bm{k},\omega)}{\left[\omega-\tilde{\xi}_{\bm{k}}-\mathrm{Re}\Sigma^{R}(\bm{k},\omega)\right]^{2}+\left[\mathrm{Im}\Sigma^{R}(\bm{k},\omega)\right]^{2}},\nonumber \\
\label{eq:spectral function}
\end{eqnarray}
and the retarded self-energy is obtained by analytic continuation
to real frequencies, 
\begin{equation}
\Sigma^{R}(\bm{k},\omega)=\Sigma(\bm{k},\omega+i0^{+}).
\end{equation}
We note that the spectral function of hard-core bosons fulfills the
following sum rule,\cite{Fauseweh14} 
\begin{equation}
\int_{-\infty}^{\infty}d\omega\,A(\bm{k},\omega)=\left\langle \left[b_{\bm{k}},b_{\bm{k}}^{\dagger}\right]\right\rangle =1-2n,
\end{equation}
with 
\begin{equation}
n=\frac{1}{N}\sum_{\bm{k}}n_{\bm{k}}=\frac{1}{N}\sum_{\bm{k}}\int_{-\infty}^{\infty}dx\,A(\bm{k},x)n_{B}(x).
\end{equation}
Our goal is to calculate the self-consistent solution for the spectral
function $A(\bm{k},\omega)$. But before we can do this, we have to
take the limit $U\to\infty$ analytically.

\subsubsection{Taking the limit $U\to\infty$}

We can write the effective interaction $\Gamma(P)$ as 
\begin{equation}
\Gamma(P)=\frac{1}{\Pi(P)}+\delta\Gamma(P),
\end{equation}
where the second term 
\begin{equation}
\delta\Gamma(P)=-\frac{1}{\Pi(P)}\frac{1}{1+U\Pi(P)}
\end{equation}
does also contribute to the $U\to\infty$ limit because the denominator
of $\delta\Gamma(P)$ can vanish at high frequencies $\omega_{p}\sim\mathcal{O}(U)$
leading to an additional delta function contribution which has to
be taken into account. This subtlety of the $U\rightarrow\infty$
limit has been noticed only quite recently by FSU.\cite{Fauseweh14}
We now follow FSU to derive the correct hard-core limit for our model.
First of all, we note that $\Pi(\bm{p},\omega)\propto1/\omega$ for
$\omega\to\infty$. This allows us to introduce the spectral representation
\begin{equation}
\Pi(\bm{p},\omega)=\int_{-\infty}^{\infty}dx\,\frac{\rho(\bm{p},x)}{\omega-x},
\end{equation}
where 
\begin{eqnarray}
\rho(\bm{p},\omega) & = & -\frac{1}{\pi}\mathrm{Im}\Pi(\bm{p},\omega+i0^{+})\nonumber \\
 & = & -\frac{1}{N}\sum_{\bm{q}}\int_{-\infty}^{\infty}dx\,A(\bm{q},x)A(\bm{p}-\bm{q},\omega-x)\nonumber \\
 &  & \times\left[n_{B}(x)-n_{B}(-x)\right].\label{eq:rho}
\end{eqnarray}
Now, we use the fact that for $\omega\to\infty$ 
\begin{equation}
\Gamma(\bm{p},\omega)-U=\frac{-U^{2}\Pi(\bm{p},\omega)}{1+U\Pi(\bm{p},\omega)}\propto\frac{1}{\omega},
\end{equation}
because $\Pi(\bm{p},\omega)\propto1/\omega$ for $\omega\to\infty$.
This implies that $\Gamma(\bm{p},\omega)-U$ has the spectral representation
\begin{equation}
\Gamma(\bm{p},\omega)-U=\int_{-\infty}^{\infty}dx\,\frac{\bar{\rho}(\bm{p},x)}{\omega-x},\label{eq:gamma spectral}
\end{equation}
where 
\begin{eqnarray}
\bar{\rho}(\bm{p},\omega) & = & -\frac{1}{\pi}\mathrm{Im}\left[\Gamma(\bm{p},\omega+i0^{+})\right]\nonumber \\
 & = & f(\bm{p},\omega)-\frac{1}{\pi}\mathrm{Im}\left[\delta\Gamma(\bm{p},\omega+i0^{+})\right],\label{eq:rho_bar}
\end{eqnarray}
with 
\begin{equation}
f(\bm{p},\omega)=\frac{-\rho(\bm{p},\omega)}{\left[\mathcal{P}\int_{-\infty}^{\infty}dx\frac{\rho(\bm{p},x)}{\omega-x}\right]^{2}+\left[\pi\rho(\bm{p},\omega)\right]^{2}}.\label{eq:f}
\end{equation}
Here, $\mathcal{P}$ denotes the Cauchy principal value which arises
from the identity $1/(\omega+i0^{+})=\mathcal{P}(1/\omega)-i\pi\delta(\omega)$.
For the contribution of $\delta\Gamma$ to $\bar{\rho}(\bm{p},\omega)$
we recall that the denominator of $\delta\Gamma(\bm{p},\omega)$ can
vanish when $\omega\sim\mathcal{O}(U)$ and only in that case there
can be a contribution from $\delta\Gamma$. Therefore, we expand $\Pi(\bm{p},\omega)$
for large frequencies $\omega\sim\mathcal{O}(U)$ (we take $U$ to
be very large but finite), 
\begin{equation}
\Pi(\bm{p},\omega)\approx\frac{\rho_{0}(\bm{p})}{\omega}+\frac{\rho_{1}(\bm{p})}{\omega^{2}}+\mathcal{O}\left(\frac{1}{\omega^{3}}\right),
\end{equation}
where\begin{subequations} 
\begin{eqnarray}
\rho_{0}(\bm{p}) & = & \int_{-\infty}^{\infty}dx\rho(\bm{p},x),\\
\rho_{1}(\bm{p}) & = & \int_{-\infty}^{\infty}dx\,x\rho(\bm{p},x).
\end{eqnarray}
\end{subequations}We find 
\begin{eqnarray}
\delta\Gamma(P) & \approx & -\frac{1}{\frac{\rho_{0}(\bm{p})}{\omega}+\frac{\rho_{1}(\bm{p})}{\omega^{2}}+\mathcal{O}\left(\frac{1}{\omega^{3}}\right)}\nonumber \\
 &  & \times\frac{1}{1+U\frac{\rho_{0}(\bm{p})}{\omega}+U\frac{\rho_{1}(\bm{p})}{\omega^{2}}+\mathcal{O}(\frac{U}{\omega^{3}})},
\end{eqnarray}
where the terms $\mathcal{O}(1/\omega^{3})$ and $\mathcal{O}(U/\omega^{3})$
vanish at the pole $\omega\sim\mathcal{O}(U)$ in the limit $U\to\infty$,
justifying the expansion to order $1/\omega^{2}$. Therefore, we have
\begin{equation}
\delta\Gamma(P)\approx-\frac{1}{\frac{\rho_{0}(\bm{p})}{\omega}+\frac{\rho_{1}(\bm{p})}{\omega^{2}}}\,\frac{\omega^{2}}{\left(\omega-\omega_{1}(\bm{p})\right)\left(\omega-\omega_{2}(\bm{p})\right)},
\end{equation}
where the poles are given by\begin{subequations}

\begin{eqnarray}
\omega_{1}(\bm{p}) & = & -\frac{U\rho_{0}(\bm{p})}{2}-\sqrt{\frac{U^{2}\rho_{0}^{2}(\bm{p})^{2}}{4}-U\rho_{1}(\bm{p})}\nonumber \\
 & \sim & -U\rho_{0}(\bm{p}),\;\;\;U\to\infty\\
\omega_{2}(\bm{p}) & = & -\frac{U\rho_{0}(\bm{p})}{2}+\sqrt{\frac{U^{2}\rho_{0}^{2}(\bm{p})}{4}-U\rho_{1}(\bm{p})}\nonumber \\
 & \sim & -\frac{\rho_{1}(\bm{p})}{\rho_{0}(\bm{p})},\;\;\;U\to\infty.
\end{eqnarray}
\end{subequations}Only the pole at $\omega_{1}\sim\mathcal{O}(U)$
is relevant for the analytic continuation in Eq.~(\ref{eq:rho_bar})
because the other pole at $\omega_{2}\sim\mathcal{O}(U^{0})$ is spurious,
since we have expanded for large frequencies $\omega\sim\mathcal{O}(U)$.
In total, we get 
\begin{equation}
\bar{\rho}(\bm{p},\omega)=f(\bm{p},\omega)-\frac{1}{\frac{\rho_{0}(\bm{p})}{\omega}+\frac{\rho_{1}(\bm{p})}{\omega^{2}}}\,\frac{\omega^{2}}{\omega-\omega_{2}(\bm{p})}\delta(\omega-\omega_{1}(\bm{p})).
\end{equation}
We can now use the spectral representation (\ref{eq:gamma spectral})
in Eq.~(\ref{eq:self-energy from effective interaction}) and take
the limit $U\to\infty$ to get the following expression for the self-energy,

\begin{eqnarray}
\Sigma(K) & = & -\frac{2}{N}\sum_{\bm{q}}\int_{-\infty}^{\infty}dx\int_{-\infty}^{\infty}dx'A(\bm{q},x')f(\bm{q}+\bm{k},x)\nonumber \\
 &  & \times\frac{n_{B}(x)-n_{B}(x')}{i\omega_{k}+x'-x}\nonumber \\
 &  & +\frac{2}{N}\sum_{\bm{q}}\int_{-\infty}^{\infty}dx\,A(\bm{q},x)n_{B}(x)\nonumber \\
 &  & \times\left[\frac{x+i\omega_{k}}{\rho_{0}(\bm{q}+\bm{k})}-\frac{\rho_{1}(\bm{q}+\bm{k})}{\rho_{0}^{2}(\bm{q}+\bm{k})}\right].
\end{eqnarray}
By analytic continuation to real frequencies, we obtain the real and
imaginary part of the retarded self-energy,\begin{subequations} 
\begin{eqnarray}
\mathrm{Re}\Sigma^{R}(\bm{k},\omega) & = & \frac{2}{N}\sum_{\bm{q}}\int_{-\infty}^{\infty}dx\,A(\bm{q},x)n_{B}(x)\nonumber \\
 &  & \times\left[\frac{x+\omega}{\rho_{0}(\bm{q}+\bm{k})}-\frac{\rho_{1}(\bm{q}+\bm{k})}{\rho_{0}^{2}(\bm{q}+\bm{k})}\right]\nonumber \\
 &  & +\mathcal{P}\int_{-\infty}^{\infty}dx\,\frac{\rho_{\Sigma}(\bm{k},x)}{\omega-x},\label{eq:re_sigma}\\
\mathrm{Im}\Sigma^{R}(\bm{k},\omega) & = & -\pi\rho_{\Sigma}(\bm{k},\omega),
\end{eqnarray}
where 
\begin{eqnarray}
\rho_{\Sigma}(\bm{k},\omega) & = & \frac{2}{N}\sum_{\bm{q}}\int_{-\infty}^{\infty}dx\,A(\bm{q},x)f(\bm{q}+\bm{k},x+\omega)\nonumber \\
 &  & \times\left[n_{B}(x)-n_{B}(x+\omega)\right].\label{eq:rho_sigma}
\end{eqnarray}
\end{subequations}To summarize, we have obtained the self-energy
in the limit $U\to\infty$ which we can calculate starting from an
initial spectral function. Via Eq.~(\ref{eq:spectral function})
we can then calculate the next iteration of the spectral function
allowing us to find a self-consistent solution for the spectral function.
After each iteration the Hartree-Fock parameters $n$, $\tau_{1}$,
$\tau_{2}$, and $\tau_{3}$ have to be updated via the self-consistency
equations (\ref{eq:tau}) and (\ref{eq:n}) using 
\begin{equation}
n_{\bm{k}}=\left\langle b_{\bm{k}}^{\dagger}b_{\bm{k}}\right\rangle =\int_{-\infty}^{\infty}dx\,A(\bm{k},x)n_{B}(x).\label{eq:n_k}
\end{equation}

\section{Numerical results for the one-dimensional $XY$ model\label{sec:numerical-Results XY}}

Before applying the above approach to the hard-core boson model for
Cs$_{2}$CuCl$_{4}$, it is instructive to test its validity for the
exactly solvable one-dimensional spin-$1/2$ $XY$ model in a magnetic
field. Although a similar model has been already studied in detail
in Ref.~{[}\onlinecite{Fauseweh14}{]}, the breakdown of the self-consistent
ladder approximation in the vicinity of the quantum critical point
has not been investigated.

The $XY$ model in one dimension is given by 
\begin{equation}
\ensuremath{\mathcal{H}_{\mathrm{1D}}=J\sum_{i}\left(S_{i}^{x}S_{i+1}^{x}+S_{i}^{y}S_{i+1}^{y}\right)-h\sum_{i}S_{i}^{z}},
\end{equation}
which can again be mapped to hard-core bosons (neglecting constant
terms),

\begin{equation}
\mathcal{H}_{\mathrm{1D}}=\sum_{k}\xi_{k}b_{k}^{\dagger}b_{k}+\frac{U}{2N}\sum_{k,k',q}b_{k+q}^{\dagger}b_{k'-q}^{\dagger}b_{k'}b_{k},
\end{equation}
with excitation energy 
\begin{equation}
\xi_{k}=J\left[\cos(k_{x}b)+1\right]-\mu,
\end{equation}
where $\mu=h_{c}-h=-\Delta$ and $h_{c}=J$. An exact solution can
be found by mapping the hard-core bosons to fermions via the Jordan-Wigner
transformation, 
\begin{equation}
b_{j}=e^{-i\pi\sum_{l<j}c_{l}^{\dagger}c_{l}}c_{j},\quad b_{j}^{\dagger}=c_{j}^{\dagger}e^{i\pi\sum_{l<j}c_{l}^{\dagger}c_{l}},
\end{equation}
resulting in the quadratic Hamiltonian 
\begin{equation}
\mathcal{H}_{\mathrm{1D}}=\sum_{k}\xi_{k}c_{k}^{\dagger}c_{k},
\end{equation}
where the operators $c_{k}^{\dagger}$ and $c_{k}$ are fermionic
creation and annihilation operators. The hard-core boson density is
therefore exactly given by 
\begin{equation}
n=\frac{1}{N}\sum_{k}\frac{1}{e^{\beta\xi_{k}}+1},\label{eq:nexact}
\end{equation}
which can be compared to the approximate solution from the self-consistent
ladder approximation for the hard-core boson model. We note that for
low temperatures at $\mu=0$, the exact density (\ref{eq:nexact})
has the following asymptotic behavior, 
\begin{equation}
n\sim\frac{\sqrt{2T/J}}{\pi}\int_{0}^{\infty}\!\!dx\,\frac{1}{e^{x^{2}}+1}\approx0.241\sqrt{T/J},
\end{equation}
in agreement with the expected behavior of one-dimensional bosons
at the quantum critical point.\cite{Sachdev94} 
\begin{figure}
\begin{centering}
\includegraphics{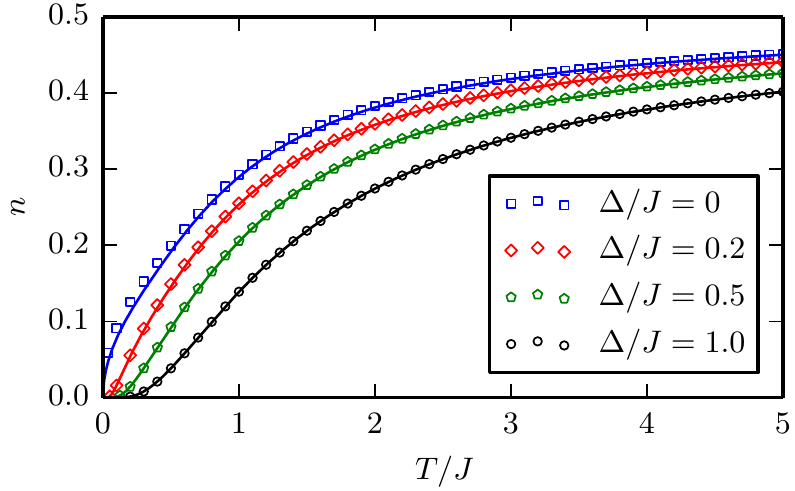} 
\par\end{centering}

\caption{(Color online) Comparison of the results for the boson density $n$
obtained from the self-consistent ladder approximation (symbols) with
the exact result (solid lines) at different energy gaps $\Delta$
for the one-dimensional $XY$ model.\label{fig:1d n}}
\end{figure}

\begin{figure}
\begin{centering}
\includegraphics{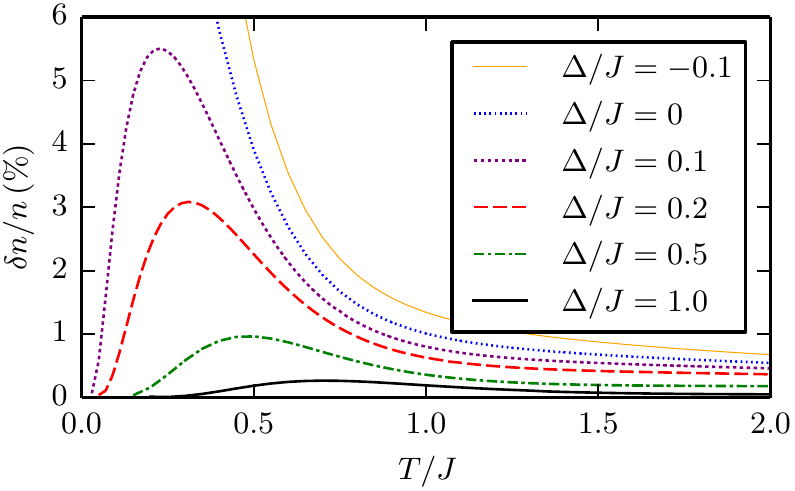} 
\par\end{centering}

\caption{(Color online) Relative error $\delta n/n$ of the boson density (see
Eq.~(\ref{eq:error})) in the self-consistent ladder approximation
for the one-dimensional $XY$ model as a function of temperature at
different energy gaps $\Delta$.\label{fig:1d delta n}}
\end{figure}

In Fig.~\ref{fig:1d n} we compare the approximate result for the
boson density $n$ obtained within the self-consistent ladder approximation
with the exact solution. The relative error $\delta n/n$ of the approximate
result is shown in Fig.~\ref{fig:1d delta n} where we define 
\begin{equation}
\delta n/n=\frac{n_{\mathrm{ladder}}-n_{\mathrm{exact}}}{n_{\mathrm{exact}}}.\label{eq:error}
\end{equation}
Here, $n_{\mathrm{ladder}}$ is the result from the self-consistent
ladder approximation and $n_{\mathrm{exact}}$ is the exact result.
For a finite gap, we see that the error vanishes for low and high
temperatures with a maximum error at $T\approx\Delta$, while the
error becomes smaller for larger energy gaps. For $\Delta=0$, the
error keeps increasing for smaller temperatures, getting closer to
the quantum critical point at $T=0$, but decreases for higher temperatures.
We conclude that the self-consistent ladder approximation gives good
results over the whole temperature range for finite energy gaps $\Delta\gtrsim0.1\;J$.

\section{Numerical results for $\mathbf{Cs_{2}CuCl_{4}}$\label{sec:numerical-Results Cs2CuCl4}}

In this section, we apply our hard-core boson approach described in
Sec.~\ref{sec:theoretical-approach} to the relevant model for Cs$_{2}$CuCl$_{4}$
given in Sec.~\ref{sec:model}. From the numerical solution of the
self-consistent ladder approximation we calculate the spectral function
of the hard-core bosons at finite temperatures for different magnetic
fields in the regime $B>B_{c}$ where the energy gap $\Delta>0$ is
finite. Given the spectral function, we can calculate the magnetization,
the internal energy, and the transverse part of the spin dynamic structure
factor. From the magnetization and internal energy we obtain the magnetic
susceptibility and the specific heat by numerical differentiation.
Finally, we compare our results with experimental data for the specific
heat.\cite{Radu05,Radu07} \setcounter{subsubsection}{0} Technical
details of the numerical solution of the self-consistent ladder approximation
can be found in Appendix~A and Appendix~B.

\subsubsection{Spectral function}

\begin{figure*}
\begin{centering}
\includegraphics{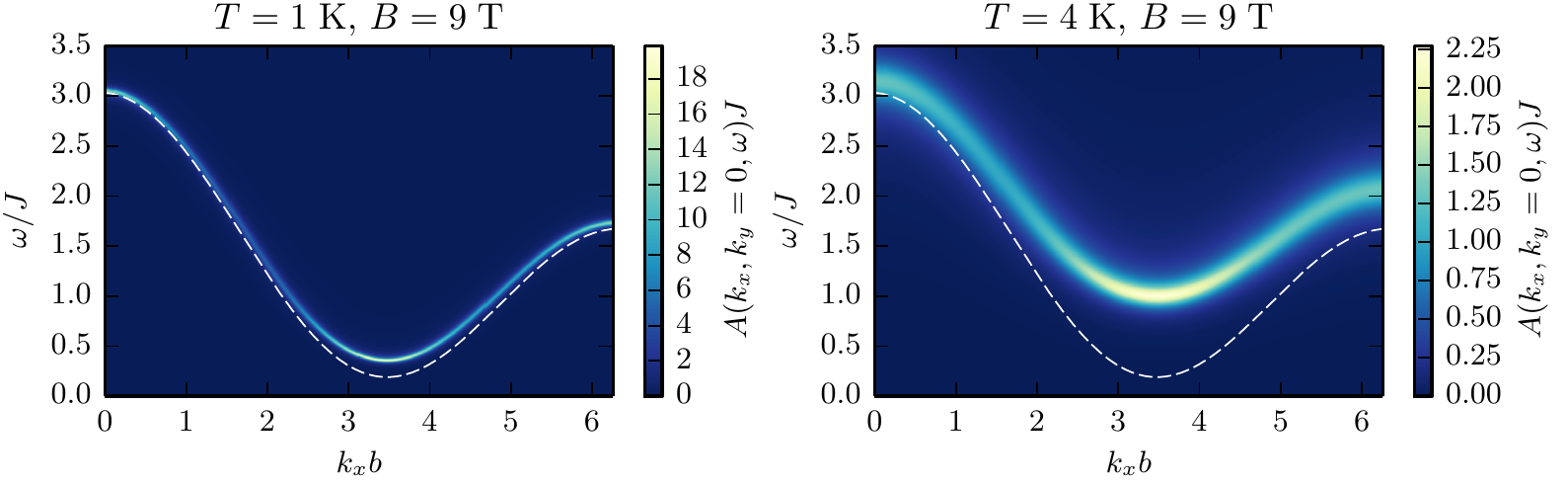}
\par\end{centering}

\caption{(Color online) Contour plots of the spectral function $A(\bm{k},\omega)$
of the hard-core bosons at $k_{y}=0$ for temperatures of $1\;\mathrm{K}$
and $4\;\mathrm{K}$ in a magnetic field $B=9\;\mathrm{T}$ corresponding
to an energy gap $\Delta=0.19J$. The white dashed line is the bare
excitation energy $\xi_{\bm{k}}$ given by Eq.~(\ref{eq:bare excitation energy}).\label{fig:contour spectral function}}
\end{figure*}

\begin{figure*}
\begin{centering}
\includegraphics{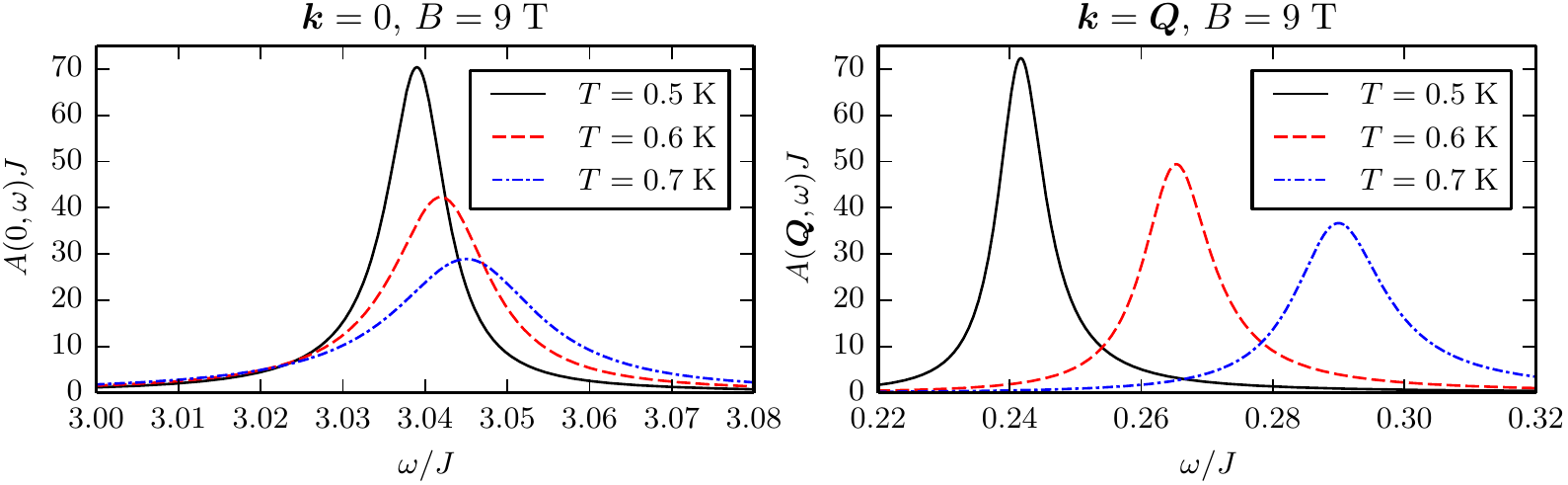} 
\par\end{centering}

\caption{(Color online) Spectral function $A(\bm{k},\omega)$ at $\bm{k}=0$
and at $\bm{k}=\bm{Q}$ (at the minimum of the dispersion) in a magnetic
field $B=9\;\mathrm{T}$ corresponding to an energy gap $\Delta\approx0.19J$.
\label{fig:spectral function min max}}
\end{figure*}

Due to the finite energy gap $\Delta>0$, at zero temperature the
spectral function is exactly given by the non-interacting spectral
function, 
\begin{equation}
A(\bm{k},\omega)=A_{0}(\bm{k},\omega)=\delta(\omega-\xi_{\bm{k}}).
\end{equation}
At finite temperatures, interactions will lead to a renormalization
of the excitation energy $\xi_{\bm{k}}$ and a broadening of the delta
peaks. Since we are treating the $J_{\bm{q}}$ interaction term on
a Hartree-Fock level, this alone would only renormalize the excitation
energy by $\xi_{\bm{k}}\to\tilde{\xi}_{\bm{k}}$ resulting in 
\begin{equation}
A(\bm{k},\omega)=\delta(\omega-\tilde{\xi}_{\bm{k}}).
\end{equation}
Taking in addition the hard-core interaction via the self-consistent
ladder approximation into account will lead to a broadening of the
spectral function with rising temperature, as shown in Fig.~\ref{fig:contour spectral function}.
Besides the broadening, we notice that the bandwidth shrinks with
rising temperature and the minimum of the spectral function gets shifted
to higher energies increasing the effective energy gap from its bare
value $\Delta$ at $T=0$. In Fig.~\ref{fig:spectral function min max}
we contrast the behavior of the spectral function at $\bm{k}=0$ and
at the minimum of the dispersion $\bm{k}=\bm{Q}$. While at $\bm{k}=0$
the position of the peak only moves to slightly higher energies, at
the minimum of the dispersion the peak gets considerably shifted to
higher energies. Due to the finite frequency resolution in our numerical
calculation (see Appendix A), we cannot reach arbitrarily low temperatures
and are restricted to temperatures $T\apprge0.2\Delta$ where the
spectral function is not too narrow to be resolved. However, in the
temperature range $T\lesssim0.2\Delta$, the hard-core interaction
can be neglected and we can then just use the self-consistent Hartree-Fock
decoupling without hard-core interaction, as we will show further
below. 

The spectral function can be related to the in-plane components of
the spin dynamic structure factor. The spin dynamic structure factor
is defined by 
\begin{equation}
S^{\alpha\beta}(\bm{k},\omega)=\int_{-\infty}^{\infty}\frac{dt}{2\pi}e^{i\omega t}\left\langle S_{-\bm{k}}^{\alpha}(t)S_{\bm{k}}^{\beta}(0)\right\rangle ,
\end{equation}
where $\alpha,\beta=x,y,z$ and the Fourier transforms of the spin
operators are defined via 
\begin{equation}
S_{\bm{k}}^{\alpha}=\frac{1}{\sqrt{N}}\sum_{i}e^{-i\bm{k}\cdot\bm{R}_{i}}S_{i}^{\alpha}.
\end{equation}
The in-plane components of the spin dynamic structure factor are given
by\begin{subequations}

\begin{eqnarray}
S^{xx}(\bm{k},\omega) & = & S^{yy}(\bm{k},\omega)\nonumber \\
 & = & \frac{1}{4}\frac{1}{1-e^{-\beta\omega}}\left(A(\bm{k},\omega)+A(-\bm{k},\omega)\right),\nonumber \\
\\
S^{xy}(\bm{k},\omega) & = & -S^{yx}(\bm{k},\omega)\nonumber \\
 & = & \frac{1}{4}\frac{1}{1-e^{-\beta\omega}}\left(A(\bm{k},\omega)-A(-\bm{k},\omega)\right),\nonumber \\
\end{eqnarray}
\end{subequations}where $S^{xy}(\bm{k},\omega)$ and $S^{yx}(\bm{k},\omega)$
do not vanish due to the broken inversion symmetry. The $U(1)$ symmetry
due to the spin-rotational invariance with respect to the $z$ axis
requires that 
\begin{equation}
S^{xz}(\bm{k},\omega)=S^{zx}(\bm{k},\omega)=S^{yz}(\bm{k},\omega)=S^{zy}(\bm{k},\omega)=0.
\end{equation}
The $S^{zz}$ component of the spin dynamic structure factor cannot
be simply expressed in terms of the spectral function because it is
a two-particle Green function in terms of the hard-core boson operators,

\begin{eqnarray}
S^{zz}(\bm{k},\omega) & = & \frac{1}{N}\sum_{\bm{q},\bm{q}'}\int_{-\infty}^{\infty}\frac{dt}{2\pi}e^{i\omega t}\nonumber \\
 &  & \times\left\langle b_{\bm{q}}^{\dagger}(t)b_{\bm{q}-\bm{k}}(t)b_{\bm{q}'}^{\dagger}(0)b_{\bm{q}'+\bm{k}}(0)\right\rangle .
\end{eqnarray}

\subsubsection{Magnetic moment and magnetic susceptibility}

The magnetic moment per site is given by 
\begin{equation}
m=\left\langle S_{i}^{z}\right\rangle =\frac{1}{2}-n,
\end{equation}
where $n$ is the boson density per site which can be expressed in
terms of the spectral function, 
\begin{equation}
n=\frac{1}{N}\sum_{\bm{k}}n_{\bm{k}},
\end{equation}
with 
\begin{equation}
n_{\bm{k}}=\left\langle b_{\bm{k}}^{\dagger}b_{\bm{k}}\right\rangle =\int_{-\infty}^{\infty}dx\,A(\bm{k},x)n_{B}(x).
\end{equation}
We define the magnetic susceptibility $\chi$ via 
\begin{equation}
\chi=\frac{dm}{dB}.
\end{equation}
In the limit $T\to0$ the boson density vanishes and the asymptotic
low-temperature behavior of the susceptibility is therefore the one
of free bosons because all interaction are frozen out, 
\begin{equation}
\chi\propto T^{\frac{d-2}{2}}e^{-\Delta/T},
\end{equation}
where $d$ is the dimensionality ($d=2$ in our case) and $\Delta$
the energy gap.

\begin{figure}
\begin{centering}
\includegraphics{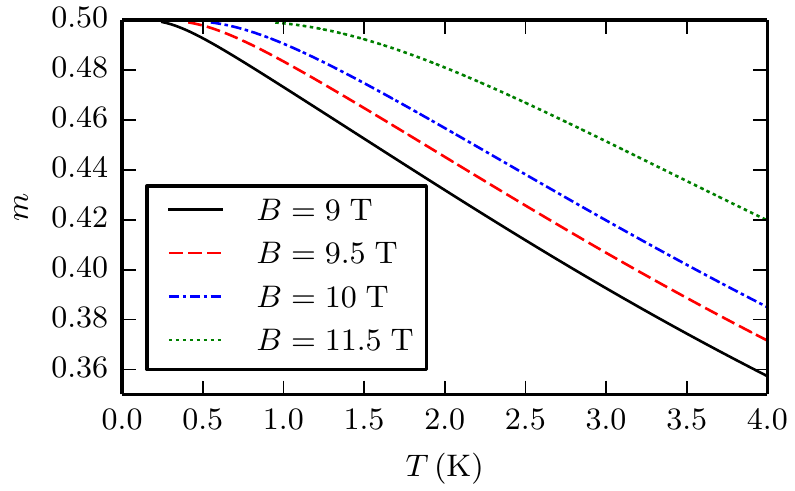} 
\par\end{centering}

\caption{(Color online) Numerical results for the magnetic moment $m$ at different
magnetic fields between $9\;\mathrm{T}$ and $11.5\;\mathrm{T}$.\label{fig:magnetization}}
\end{figure}

\begin{figure}
\begin{centering}
\includegraphics{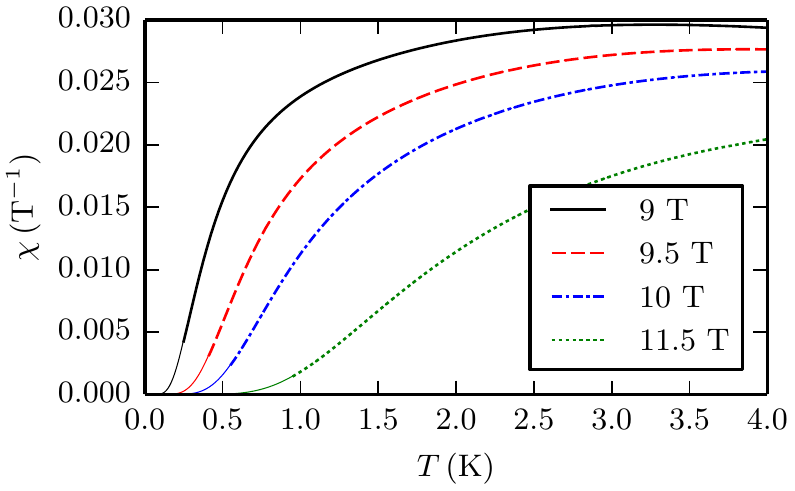} 
\par\end{centering}

\caption{(Color online) Numerical results for the magnetic susceptibility $\chi$
at different magnetic fields between $9\;\mathrm{T}$ and $11.5\;\mathrm{T}$.
The thin solid lines are low-temperature results from the Hartree-Fock
approximation without hard-core interaction, which allows us to get
results in the low-temperature regime where the self-consistent ladder
approximation cannot be used due to the limited frequency resolution.\label{fig:chi}}
\end{figure}

\begin{figure}
\begin{centering}
\includegraphics{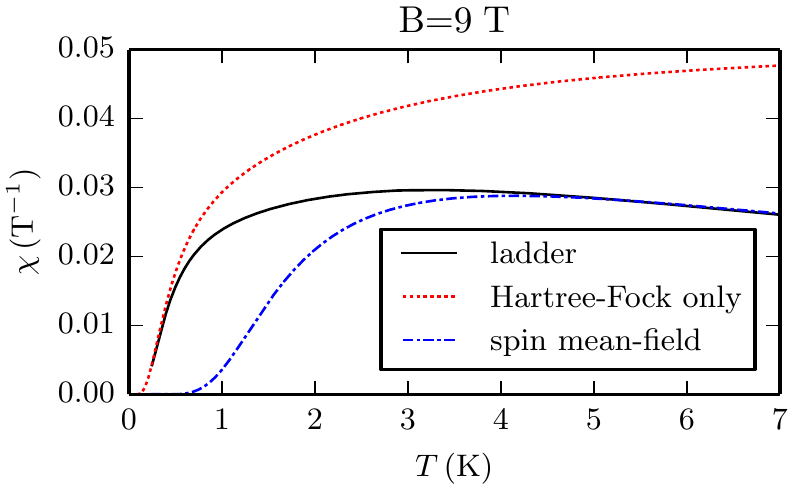} 
\par\end{centering}

\caption{(Color online) Comparison of the results for the magnetic susceptibility
$\chi$ from the self-consistent ladder approximation, Hartree-Fock
approximation without hard-core interaction, and spin mean-field theory
for a magnetic field $B=9\;\mathrm{T}$ corresponding to an energy
gap $\Delta\approx0.19J$.\label{fig:chi comparison}}
\end{figure}

The numerical results for magnetic moment and magnetic susceptibility
for different magnetic fields above the saturation field are shown
in Fig.~\ref{fig:magnetization} and Fig.~\ref{fig:chi}, respectively.
In Fig.~\ref{fig:chi comparison}, we compare our numerical results
from the self-consistent ladder approximation with the low-temperature
Hartree-Fock approximation without hard-core interaction and with
a simple spin mean-field theory, which we describe in Appendix C.
We see that for low temperatures $T\ll J$ the Hartree-Fock and the
ladder approximation give essentially the same results. This allows
us to use the Hartree-Fock approximation in the low-temperature regime
where the self-consistent ladder approximation is difficult to implement
due to the limited frequency resolution. At higher temperatures the
hard-core interaction becomes important and the high-temperature behavior
is approximately captured by the spin mean-field theory.

\subsubsection{Internal energy and specific heat}

\begin{figure}
\begin{centering}
\includegraphics{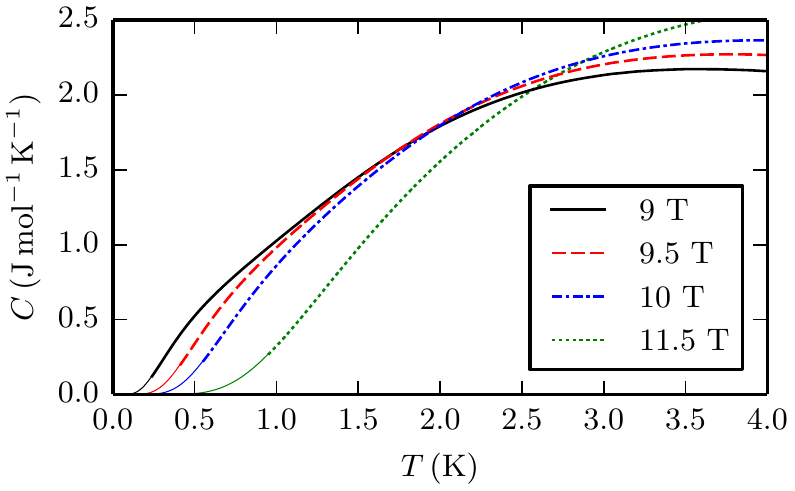} 
\par\end{centering}

\caption{(Color online) Numerical results for the specific heat $C$ at different
magnetic fields between $9\;\mathrm{T}$ and $11.5\;\mathrm{T}$.
The thin solid lines are low-temperature results from the Hartree-Fock
approximation without hard-core interaction, which allows us to get
results in the low-temperature regime where the self-consistent ladder
approximation cannot be used due to the limited frequency resolution.\label{fig:heat}}
\end{figure}

\begin{figure}
\begin{centering}
\includegraphics{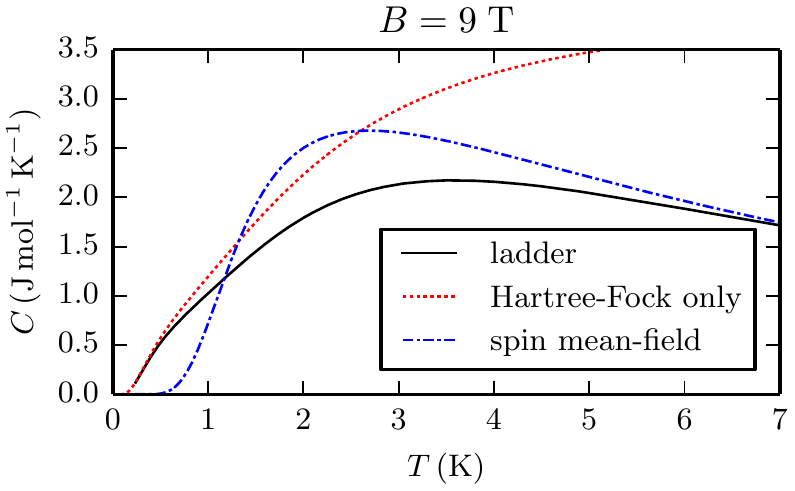} 
\par\end{centering}

\caption{(Color online) Comparison of the results for the specific heat $C$
from the self-consistent ladder approximation, Hartree-Fock approximation
without hard-core interaction, and spin mean-field theory for a magnetic
field $B=9\;\mathrm{T}$ corresponding to an energy gap $\Delta\approx0.19J$.\label{fig:heat comparison}}
\end{figure}

\begin{figure}
\begin{centering}
\includegraphics{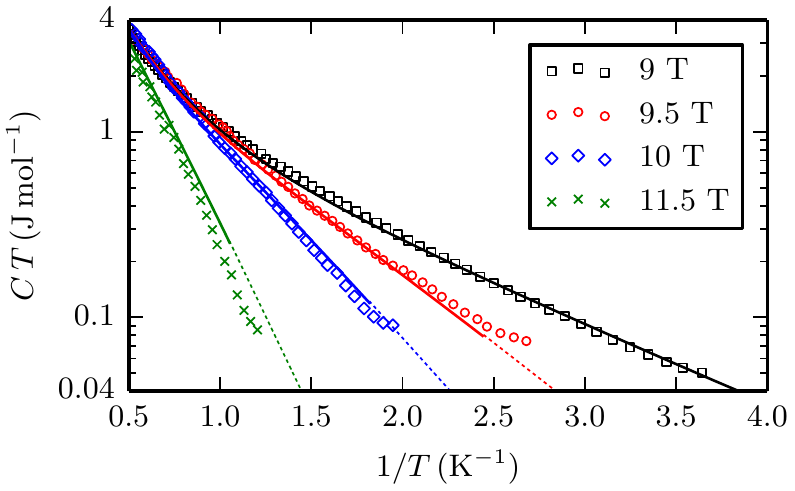}
\par\end{centering}

\caption{(Color online) Comparison of our numerical results for the specific
heat (solid lines) with experimental data from Ref.~{[}\onlinecite{Radu07}{]}
(symbols) for different magnetic fields between $9\;\mathrm{T}$ and
$11.5\;\mathrm{T}$. The dashed lines are low-temperature results
from the Hartree-Fock approximation without hard-core interaction,
which allows us to get results in the low-temperature regime where
the self-consistent ladder approximation cannot be used due to the
limited frequency resolution.\label{fig:heat experiment}}
\end{figure}

The internal energy is given by 
\begin{equation}
E=\left\langle \mathcal{H}\right\rangle =\sum_{\bm{k}}\tilde{\xi}_{\bm{k}}n_{\bm{k}}+E_{\mathrm{MF}},
\end{equation}
where the infinite on-site interaction does not contribute because
its expectation value is zero if the hard-core constraint is fulfilled.
The specific heat at constant volume is obtained by taking the temperature
derivative of the internal energy, 
\begin{equation}
C=\frac{dE}{dT}.
\end{equation}
We note that $E_{\mathrm{MF}}$ depends on temperature and has to
be taken into account for calculating the specific heat. The asymptotic
low-temperature behavior of the specific heat is given by 
\begin{equation}
C\propto T^{\frac{d-4}{2}}e^{-\Delta/T}.\label{eq:asymptotic heat}
\end{equation}
The numerical results for the specific heat at different magnetic
fields above the saturation field are shown in Fig.~\ref{fig:heat}.
In Fig.~\ref{fig:heat comparison}, we again compare our numerical
results from the self-consistent ladder approximation with the low-temperature
Hartree-Fock approximation without hard-core interaction and with
a simple spin mean-field theory described in Appendix~C. The magnetic
contribution to the specific heat of Cs$_{2}$CuCl$_{4}$ has been
measured experimentally.\cite{Radu05,Radu07} The more recent data
published in Ref.~{[}\onlinecite{Radu07}{]} differ slightly from
Ref.~{[}\onlinecite{Radu05}{]} for $B=11.5\;\mathrm{T}$ and are
in better agreement with the expected size of the energy gap at that
field strength. Therefore, in Fig.~\ref{fig:heat experiment} we
compare our results with the experimental data from Ref.~{[}\onlinecite{Radu07}{]},
where we find that our theory captures the experimentally observed
behavior both qualitatively and quantitatively. At low temperatures,
the slope in the logarithmic plot of $CT$ versus $1/T$ in Fig.~\ref{fig:heat experiment}
is given by $-\Delta$, which follows directly from Eq.~(\ref{eq:asymptotic heat}).

\section{Summary and conclusions\label{sec:summary}}

We have mapped the spin-1/2 Heisenberg model describing Cs$_{2}$CuCl$_{4}$
to a model of hard-core bosons where the hard-core constraint has
been taken into account by an infinite on-site repulsion. Since we
have only considered magnetic fields $B>B_{c}$ (along the $a$ axis
perpendicular to the lattice plane), we had to deal with gapped hard-core
bosons. Due to the energy gap, the hard-core interaction can be taken
into account using the self-consistent ladder approximation\cite{Fauseweh14}
and the remaining interactions can be treated within the self-consistent
Hartree-Fock approximation. Before applying this method to Cs$_{2}$CuCl$_{4}$,
we have investigated for the exactly solvable one-dimensional $XY$
model how the ladder approximation breaks down in the vicinity of
the critical field $B_{c}$, finding that the ladder approximation
for finite energy gaps $\Delta$ works well both at low and high temperatures
and the deviations, maximal at $T\approx\Delta$, decrease with rising
energy gap $\Delta$. We have then calculated the spectral function
of the hard-core bosons for Cs$_{2}$CuCl$_{4}$ from which we have
obtained the magnetic susceptibility and the specific heat. The calculated
specific heat is in good agreement with the available experimental
data. We conclude that the self-consistent ladder approximation in
combination with a self-consistent Hartree-Fock decoupling of the
non-hard-core interactions gives an accurate description of the physical
properties of gapped hard-core bosons in two dimensions at finite
temperatures. An extension to three dimensions is straightforward
and would only increase the numerical effort due to an increasing
number of lattice sites. Our methods can also be directly applied
to the material class Cs$_{2}$Cu(Cl$_{4-x}$Br$_{x}$), where chlorine
is partially substituted by bromine which changes the strength of
the exchange couplings and the ratio $J'/J$.\cite{Krueger10,Cong11,Foyevtsova11}
While in our work we started from a spin-1/2 model which we mapped
to hard-core bosons, our theoretical approach is applicable whenever
the elementary excitations can be described by gapped hard-core bosons;
some examples are discussed in Ref.~{[}\onlinecite{Fauseweh14}{]}.
In the case of $S>1/2$, a mapping to hard-core bosons is not known,
but mapping the spins to canonical bosons is possible by using, for
example, the Holstein-Primakoff transformation,\cite{Holstein40}
where the constraint on the boson occupation number per site, $\hat{n}_{i}\leq2S$,
is difficult to take into account analytically. Therefore, spin-wave
theories based on such a mapping to canonical bosons are only valid
for low boson densities at low temperatures, where the
constraint is not important and the usual expansion in $\hat{n}_{i}/S$
is justified.
\begin{acknowledgments}
We thank Michael Lang, Lars Postulka, and Bernd Wolf for useful discussions.
Financial support by the DFG via SFB/TRR49 is gratefully acknowledged.

\appendix
\end{acknowledgments}

\section*{Appendix A: Numerical details}

\global\long\def\theequation{A.\arabic{equation}}
In this appendix, we give more details on the numerical solution of
the self-consistency equations for the spectral function $A(\bm{k},\omega)$.
To find a self-consistent solution for $A(\bm{k},\omega)$, we have
to start from an initial spectral function $A_{\mathrm{init}}(\bm{k},\omega)$.
If we would just have the standard non-self-consistent ladder approximation,
we would replace $A(\bm{k},\omega)$ by the non-interacting spectral
function $A_{0}(\bm{k},\omega)$ and then directly calculate the spectral
function $A(\bm{k},\omega)$. It is therefore sensible to use the
non-interacting spectral function as the initial spectral function,
which is given by 
\begin{equation}
A_{0}(\bm{k},\omega)=-\frac{1}{\pi}\mathrm{Im}G_{0}(\bm{k},\omega+i0^{+})=\delta(\omega-\xi_{\bm{k}}).
\end{equation}
We note here that at $T=0$ the spectral function is not affected
by interactions, $A(\bm{k},\omega)=A_{0}(\bm{k},\omega)$, because
the ground state is the vacuum without any bosons due to the energy
gap. For the numerical calculation, we replace the delta function
by a box function of finite width $\eta$ (e.g., $\eta=0.1\,J$) centered
at $\omega=\xi_{\bm{k}}$. This is fine as long as $\xi_{\bm{k}}>0$
which is the case for magnetic fields $B>B_{c}$. For $\xi_{\bm{k}}\leq0$,
we have to take the sign of the spectral function into account,\cite{Negele+Orland}
\begin{equation}
\mathrm{sgn}\left(A(\bm{k},\omega)\right)=\mathrm{sgn}\,\omega.
\end{equation}
Therefore, a positive delta peak is not permitted for negative frequencies
and the non-interacting spectral function cannot be used for values
of $\bm{k}$ with $\xi_{\bm{k}}\leq0$. In our calculations, we instead
place a step function at a small positive frequency when $\xi_{\bm{k}}\leq0$.
This allows us to find a self-consistent solution even for $B\leq B_{c}$.

Having chosen an initial spectral function, the next step is to calculate
$\rho(\bm{p},\omega)$ via Eq.~(\ref{eq:rho}), which is a multi-dimensional
convolution that can be calculated with the fast Fourier transform
method, e.g., using the FFTW library.\cite{Frigo05} Then $f(\bm{p},\omega)$
can be obtained from Eq.~(\ref{eq:f}), where the principal value
integral can also be evaluated as a convolution.\cite{Liu81} Next,
the calculation of $\rho_{\Sigma}(\bm{k},\omega)$ via Eq.~(\ref{eq:rho_sigma})
and $\mathrm{Re}\Sigma^{R}(\bm{k},\omega)$ via Eq.~(\ref{eq:re_sigma})
also involves convolutions. While the values of $\bm{k}$ are naturally
discretized for a finite lattice, as discussed in Appendix B, the
real frequencies $\omega$ have to be artificially discretized, leading
to a limited frequency resolution, and a frequency cutoff has to be
introduced (e.g., $|\omega|<20\,J$). When using the fast Fourier
transform method to evaluate the convolutions, this treats the functions
as periodic both in momentum and frequency space, leading to a wrap-around
effect in the frequency dependence of the calculated functions. This
wrap-around error can be dealt with by setting the spectral function
$A(\bm{k},\omega)$ to zero for frequencies larger than a certain
cutoff (e.g., for $|\omega|>10\,J$). In our calculations, we typically
used lattice sizes up to $4096$ sites and up to $131\,072$ frequency
points.

To achieve convergence, a simple mixing update procedure has to be
used, where the updated spectral function and Hartree-Fock parameters
are set to be a mixture of the previous iteration and the new values
from the self-consistency equations. In our case, a mixing of $50\%$
worked well. We note that in the case without the self-consistent
Hartree-Fock decoupling (e.g., for an $XY$ model), mixing is not
necessary to achieve convergence. The converged numerical result should
(up to a small numerical error) fulfill the sum rule \cite{Fauseweh14}
\begin{equation}
\int_{-\infty}^{\infty}d\omega\,A(\bm{k},\omega)=1-2n.
\end{equation}

\section*{Appendix B: Brillouin zone discretization}

\global\long\def\theequation{B.\arabic{equation}}
\setcounter{equation}{0}The use of fast Fourier transform methods
is based on the periodicity of the transformed functions. Therefore,
the Brillouin zone should not be arbitrarily discretized because that
would in most cases destroy the periodicity. Still, there is an infinite
number of possible parameterizations of the Brillouin zone. In our
work, we have used two parameterizations which we will present here.
The first parameterization starts from the lattice basis

\begin{equation}
\bm{a}_{1}=b\hat{\bm{x}},\;\;\;\bm{a}_{2}=-\frac{b}{2}\hat{\bm{x}}+\frac{c}{2}\hat{\bm{y}},
\end{equation}
with the corresponding reciprocal basis

\begin{equation}
\bm{b}_{1}=\frac{2\pi}{b}\hat{\bm{x}}+\frac{2\pi}{c}\hat{\bm{y}},\;\;\;\bm{b}_{2}=\frac{4\pi}{c}\hat{\bm{y}}.
\end{equation}
The lattice momentum vectors can then be expanded in terms of the
reciprocal basis, 
\begin{equation}
\bm{k}=k_{1}\bm{b}_{1}+k_{2}\bm{b}_{2},
\end{equation}
where the periodic boundary conditions dictate that\begin{subequations}
\begin{eqnarray}
k_{1}=\frac{l_{1}}{N_{1}},\;\;\; & l_{1}\in\left\{ 0,...,N_{1}-1\right\} ,\\
k_{2}=\frac{l_{2}}{N_{2}},\;\;\; & l_{2}\in\left\{ 0,...,N_{2}-1\right\} .
\end{eqnarray}
\end{subequations}The total number of lattice sites is $N=N_{1}N_{2}$.
To obtain a uniform mesh,\cite{Moreno92} we have to choose 
\begin{equation}
N_{2}=2N_{1}.
\end{equation}
The second (primed) parameterization starts from the lattice basis

\begin{equation}
\bm{a}_{1}'=\frac{b}{2}\hat{\bm{x}}-\frac{c}{2}\hat{\bm{y}},\;\;\;\bm{a}_{2}'=\frac{b}{2}\hat{\bm{x}}+\frac{c}{2}\hat{\bm{y}},
\end{equation}
with the corresponding reciprocal basis

\begin{equation}
\bm{b}'_{1}=\frac{2\pi}{b}\hat{\bm{x}}-\frac{2\pi}{c}\hat{\bm{y}},\;\;\;\bm{b}'_{2}=\frac{2\pi}{b}\hat{\bm{x}}+\frac{2\pi}{c}\hat{\bm{y}}.
\end{equation}
The lattice momentum vectors can again be expanded in terms of the
reciprocal basis, 
\begin{equation}
\bm{k}=k_{1}'\bm{b}_{1}'+k_{2}'\bm{b}_{2}',
\end{equation}
where the periodic boundary conditions dictate that\begin{subequations}
\begin{eqnarray}
k_{1}'=\frac{l_{1}'}{N_{1}'},\;\;\; & l_{1}'\in\left\{ 0,...,N_{1}'-1\right\} ,\\
k_{2}'=\frac{l_{2}'}{N_{2}'},\;\;\; & l_{2}'\in\left\{ 0,...,N_{2}'-1\right\} .
\end{eqnarray}
\end{subequations}The total number of lattice sites is $N=N_{1}'N_{2}'$.
To obtain a uniform mesh,\cite{Moreno92} we have to choose 
\begin{equation}
N_{2}'=N_{1}'.
\end{equation}

\section*{Appendix C: Spin mean-field approximation}

\global\long\def\theequation{C.\arabic{equation}}
\setcounter{equation}{0}We expect that at high temperatures $T\gg J$,
the spins decouple and it is sufficient to describe the spin-spin
interactions on a mean-field level where the effects of the interactions
are approximated by an effective magnetic field. To derive this mean-field
description, we start from the Hamiltonian (\ref{eq:hamiltonian}),

\begin{equation}
\ensuremath{\mathcal{H}=\frac{1}{2}\sum_{ij}\left[J_{ij}\bm{S}_{i}\cdot\bm{S}_{j}+\bm{D}_{ij}\cdot(\bm{S}_{i}\times\bm{S}_{j})\right]-h\sum_{i}S_{i}^{z}}.
\end{equation}
First, we note that only the $z$-component of the expectation values
of the spin operators does not vanish, 
\begin{equation}
\left\langle \bm{S}_{i}\right\rangle =m\hat{\bm{z}},\;\;\;m=\left\langle S_{i}^{z}\right\rangle .
\end{equation}
Expanding up to linear order in fluctuations from this expectation
value, we find 
\begin{equation}
\mathcal{H}\approx-NJ_{0}\frac{m^{2}}{2}-h_{\mathrm{eff}}\sum_{i}S_{i}^{z},
\end{equation}
where the effective magnetic field is given by 
\begin{equation}
h_{\mathrm{eff}}=h-J_{0}m,
\end{equation}
with 
\begin{equation}
J_{0}=2J+4J'.
\end{equation}
The magnetic moment $m$ is obtained by solving the self-consistency
equation 
\begin{equation}
m=\frac{1}{2}\tanh\left(\frac{\beta}{2}h_{\mathrm{eff}}\right),
\end{equation}
and the energy in this mean-field approximation is a simple function
of magnetic field and magnetic moment, 
\begin{equation}
E=N\left(\frac{1}{2}J_{0}m^{2}-mh\right).
\end{equation}

 \bibliographystyle{apsrev4-1}

\end{document}